\documentclass[journal]{IEEEtran}

\usepackage{graphicx}
\graphicspath{{}}  
\usepackage{enumerate}
\usepackage{xcolor}
\usepackage{booktabs}
\usepackage{float}
\usepackage{cite}
\usepackage[cmex10]{amsmath}
\usepackage{bm}
\usepackage[normalem]{ulem}  

\begin{document}

\title{One-Port Direct/Reverse Method for Characterizing VNA Calibration Standards }

\author{Raul~A.~Monsalve,~\IEEEmembership{Member,~IEEE,}
        Alan~E.~E.~Rogers,~\IEEEmembership{Life Member,~IEEE,}
        Thomas~J.~Mozdzen,~\IEEEmembership{Student Member,~IEEE,}
        and~Judd~D.~Bowman,~\IEEEmembership{Member,~IEEE}
\thanks{Manuscript received January XX, 2015; revised January YY, 2015. This work was supported by the NSF through research awards for the Experiment to Detect the Global EoR Signature (AST-0905990 and AST-1207761) and by NASA through Cooperative Agreements for the Lunar University Network for Astrophysics (NNA09DB30A) and the Nancy Grace Roman Technology Fellowship (NNX12AI17G).}
\thanks{R. A. Monsalve, T. J. Mozdzen, and J. D. Bowman are with the School of Earth and Space Exploration, Arizona State University, Tempe, AZ 85287, USA. e-mail: raul.monsalve@asu.edu.}
\thanks{A. E. E. Rogers is with the Haystack Observatory, Massachusetts Institute of Technology, Westford, MA 01886, USA.}}

\markboth{IEEE Transactions on Microwave Theory and Techniques,~Vol.~XX, No.~XX, January~20XX}%
{Shell \MakeLowercase{\textit{et al.}}: Bare Demo of IEEEtran.cls for Journals}

\maketitle

\begin{abstract}
This paper introduces a one-port method for estimating model parameters of VNA calibration standards. The method involves measuring the standards through an asymmetrical passive network connected in direct mode and then in reverse mode, and using these measurements to compute the S-parameters of the network. The free parameters of the calibration standards are estimated by minimizing a figure of merit based on the expected equality of the S-parameters of the network when used in direct and reverse modes. The capabilities of the method are demonstrated through simulations, and real measurements are used to estimate the actual offset delay of a 50-$\mathbf{\Omega}$ calibration load that is assigned zero delay by the manufacturer. The estimated delay is $38.8$ ps with a $1\sigma$ uncertainty of $2.1$ ps for this particular load. This result is verified through measurements of a terminated airline. The measurements agree better with theoretical models of the airline when the reference plane is calibrated using the new estimate for the load delay.
\end{abstract}


\IEEEpeerreviewmaketitle

\section{Introduction}
\IEEEPARstart{T}{he} search for higher accuracy in measurements of S-parameters using a vector network analyzer (VNA) has driven the development of ingenuous techniques that aim at simplifying the process of calibration and improving the modeling of calibration standards. In particular, the precise and accurate modeling of standards is an active area of research because characterization based on their physical dimensions and composition is possible only in a limited number of cases \cite{wong2, kirby, walker, eio}.

Widely used models for coaxial short-open-load-thru (SOLT) standards are presented in \cite{degroot} and \cite{scott}, which correspond to approximations to full transmission line theory \cite{ramo, marks}. The models for the open, short, and load incorporate parameters that characterize their termination elements (capacitance, inductance, and resistance, respectively) as well as their transmission line sections, or offsets (characteristic impedance, delay, and loss). The model for the thru is parameterized by its characteristic impedance, delay, and loss.

The traditional SOLT two-port VNA calibration requires precision knowledge of those four standards to solve for the correction coefficients of the $12$-term error model \cite{rytting}. The \emph{unknown thru} technique relaxes this requirement by replacing the precision thru with a generic reciprocal passive network \cite{ferrero1, ferrero2, na}. Calibration is achieved by taking advantage of the reciprocity property of the network ($S_{12}$ $=$ $S_{21}$). This technique is very useful in situations where the traditional SOLT calibration is limited by physical constraints, such as in wafer probe stations or custom test fixtures, where it is difficult to  connect a thru between the two ports. Although the passive network does not need to be known with precision, the phase of its $S_{21}$ has to be known to within a quarter of a wavelength \cite{wong1, blackham}.

A technique introduced in \cite{scott} aims at estimating parameters of the SOLT standards by measuring an asymmetrical ($S_{11}$ $\neq$ $S_{22}$) reciprocal passive network between the two VNA ports, in addition to the standards themselves. The technique solves for the free parameters by minimizing a figure of merit based on the expected reciprocity of the network. Another version of the two-port \emph{reciprocal} method, presented in \cite{jamneala1} and \cite{jamneala2}, focuses on estimating parameters of the SOL reflection standards only. The thru is characterized separately using a series of independent measurements, and the DC resistance of the $50$-$\Omega$ load is measured with a precision ohmmeter as suggested in \cite{blackham} and \cite{ridler}. A different type of method, introduced in \cite{roberts}, improves the characterization of the SOL standards by using a precision airline, which is connected to the calibrated measurement port and terminated with an offset short and a mismatch load. Ripples observed when connecting the airline are mainly due to residual source match and directivity resulting from assuming incorrect values for the calibration SOL parameters. A better set of values is obtained by iteratively minimizing the ripples.

This paper introduces the one-port direct/reverse (D/R) method for characterization of the SOL standards. Its most important feature relative to the \emph{reciprocal} approaches described above is that it only requires one-port measurements and, therefore, it is not affected by systematic effects occurring in multi-port setups. In addition, it does not rely on external reference or transfer standards that need to be characterized independently with high precision. The D/R method involves measuring the SOL standards at the reference plane, then measuring them at the end of an asymmetrical  passive network connected in direct mode, and then measuring them at the end of the network connected in reverse mode (physically reversed). This results in a total of nine measurements. In principle, several parameters could be estimated simultaneously but to keep their precision from degrading significantly it is preferable for the number of free parameters to remain low.

This work has been conducted in the context of high-accuracy reflection measurements of antennas for radio astronomy in the VHF range \cite{bowman, rogers}, and therefore the D/R method is demonstrated at frequencies up to $1$ GHz. Nonetheless, it is directly applicable at other frequencies with limitations specific to each implementation.

As a means of demonstration, the D/R method is used in this paper to estimate the offset delay of the $50$-$\Omega$ load from a Keysight (previously Agilent) 85033E $3.5$-mm calibration kit, which has a nominal value of $0$ ps. Companies usually provide realistic estimates for the parameters of the open and short but often assume that the load represents a perfect $50$-$\Omega$ termination producing no reflections, which would make the delay of its transmission line irrelevant. This is an approximation and, for some applications, inaccuracies in this parameter have a significant impact on S-parameter measurements.

\begin{figure}[t]
\centering
\includegraphics[width=0.48\textwidth]{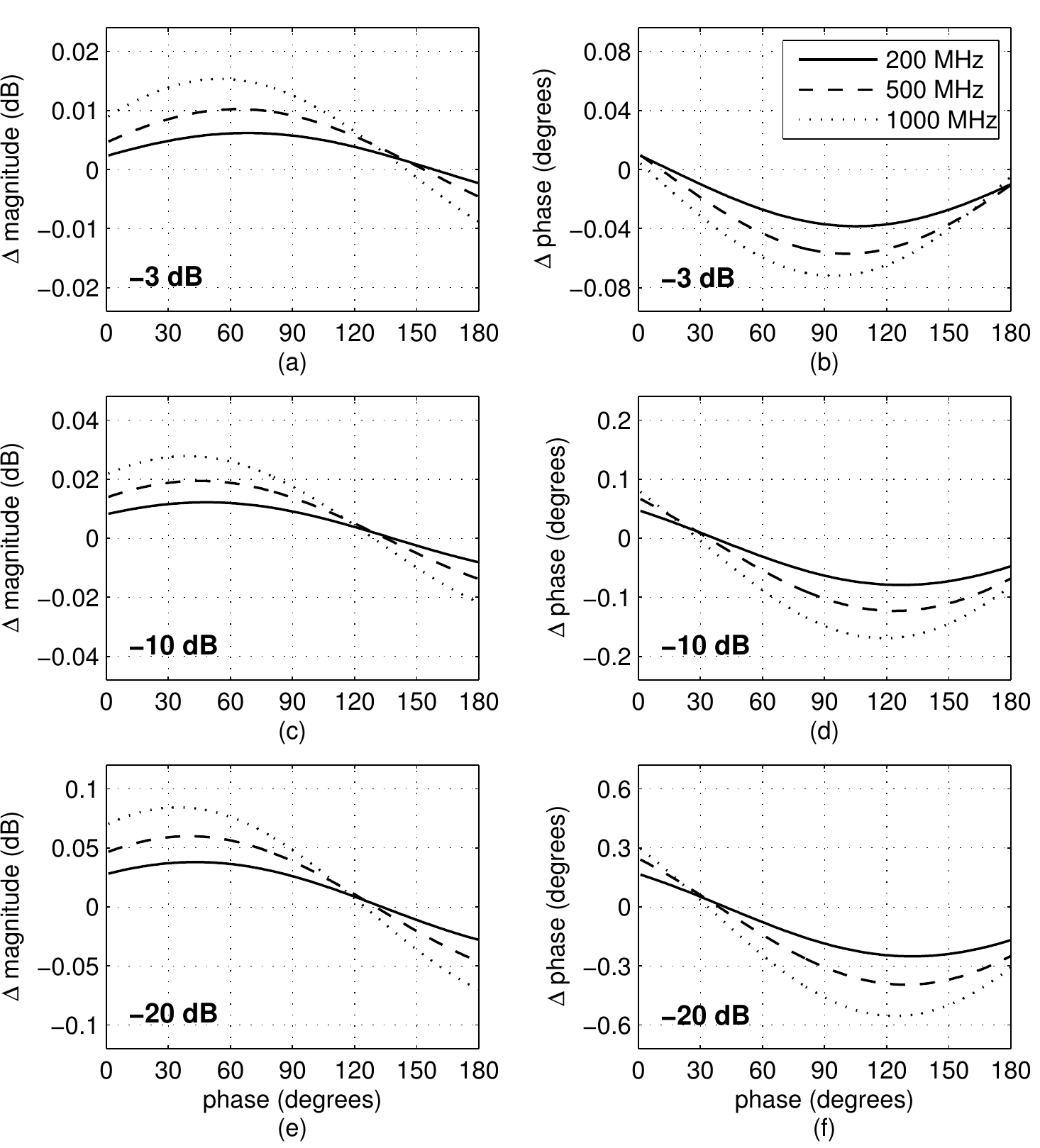}
\caption{Effect on the reflection coefficient of different DUTs, of assuming an offset delay of $0$ ps for the $50$-$\Omega$ calibration standard when its true value is $30$ ps. These results were obtained through simulations. The observed error, expressed as a difference relative to the known true reflection, depends on frequency and on the reference reflection. The left and right columns of the figure represent magnitude and phase errors, respectively. From top to bottom, the panel rows represent reference magnitudes of $-3$ dB, $-10$ dB, and $-20$ dB. The reference phase is represented along the horizontal axes and goes between $0^{\circ}$ and $180^{\circ}$.}
\label{figure_errors_delay}
\end{figure}

Fig. \ref{figure_errors_delay} shows the isolated effect of a realistic error in the load delay, on measurements of reflection coefficient. If the reference plane is calibrated with the SOL standards but assuming that the load has a delay of $0$ ps when its true value is $30$ ps, the error in the magnitude and phase of the device under test (DUT) depends on its reflection and on frequency. As an example, for a nominal reflection of $-10$ dB and $90^{\circ}$ the error at $200$ MHz is $0.01$ dB in magnitude and $-0.06^{\circ}$ in phase, which increases to $0.02$ dB and $-0.15^{\circ}$ at $1000$ MHz. A value of $30$ ps is used in this exercise because it is close to the delays reported by Keysight for the open and short of the same calibration kit.

The D/R method is described in section \ref{section_method} and demonstrated through simulations in section \ref{section_simulations}. Section \ref{section_measurements} details the parameter estimation from real measurements, section \ref{section_verification} describes the verification of the estimation and, finally, the conclusions are presented in section \ref{section_conclusion}.

\section{Method}
\label{section_method}

When a DUT is measured at the end of a two-port network, the reflection coefficient at the input of this network is given by 

\begin{equation}
\Gamma' = S_{11} + \frac{S_{12}S_{21}\Gamma}{1-S_{22}\Gamma},
\label{equation_gamma_shifted}
\end{equation}

\noindent where $\Gamma$ is the intrinsic reflection coefficient of the DUT relative to the reference impedance (usually $50$ $\Omega$), $\Gamma'$ is the reflection coefficient at the reference plane, and $S_{11}$, $S_{12}$, $S_{21}$, and $S_{22}$ are the S-parameters of the two-port network. If the S-parameters are known, $\Gamma$ can be recovered from the measurement by just inverting the equation,

\begin{equation}
\Gamma = \frac{\Gamma' - S_{11}}{S_{12}S_{21}+S_{22}(\Gamma'-S_{11})}.
\label{equation_gamma_actual}
\end{equation}

The S-parameters of the network can be computed by measuring the open, short, and load at its port 2, and then solving (\ref{equation_gamma_shifted}) in matrix form,

\begin{equation}
\begin{bmatrix}
S_{11} \\
S_{12}S_{21}-S_{11}S_{22} \\
S_{22}  
\end{bmatrix} =
\begin{bmatrix}
1 & \Gamma_\text{O} &  \Gamma_\text{O} \cdot \Gamma'_\text{O}\\
1 & \Gamma_\text{S} &  \Gamma_\text{S} \cdot \Gamma'_\text{S}\\
1 & \Gamma_\text{L} &  \Gamma_\text{L} \cdot \Gamma'_\text{L} 
\end{bmatrix}^{-1}
\begin{bmatrix}
\Gamma'_\text{O} \\
\Gamma'_\text{S} \\
\Gamma'_\text{L} 
\end{bmatrix},
\label{equation_s_parameters}
\end{equation}

\noindent where $\Gamma_{\text{O}}$, $\Gamma_{\text{S}}$, and $\Gamma_{\text{L}}$ are the reflections of the standards assumed as true, and $\Gamma'_{\text{O}}$, $\Gamma'_{\text{S}}$, and $\Gamma'_{\text{L}}$ are their values as viewed at port 1 of the network.

In this representation, ports 1 and 2 are intrinsic to the network. In other words, in direct mode port 1 is facing the measurement plane and port 2 is facing the DUT, while in reverse mode port 2 is facing the measurement plane and port 1 is facing the DUT. 

For a passive two-port network, and under ideal conditions of repeatability and linearity, the S-parameters computed in direct and reverse modes should be identical as long as the reflections from the standards assumed as true ($\Gamma_{\text{O}}$, $\Gamma_{\text{S}}$, and $\Gamma_{\text{L}}$ in (\ref{equation_s_parameters})) are correct. If this is not the case, the S-parameters recovered in direct and reverse modes will differ. These properties of passive networks can be used in principle to solve for the model parameters of the reflection standards that minimize the difference between S-parameters in direct and reverse mode.

An adequate figure of merit (FoM) has to be defined to effectively constrain the free parameters through minimization. The one used in this implementation is

\newcommand{\ra}[1]{\renewcommand{\arraystretch}{#1}}

\begin{table}\centering
\ra{1.3}
\caption{Nomenclature for the Direct/Reverse Method}
\label{table_nomenclature}
\begin{tabular}{@{}ll@{}}
\toprule
Variable    & Description \\
\midrule
$\mathbf{p}$			 		& vector of parameter estimates \\
$\mathbf{\Gamma}_\text{M}$                      & reflection of standards from model \\
$\mathbf{\Gamma}'_\text{RP}$     		& reflection of standards measured at reference plane \\
$\mathbf{\Gamma}'_\text{D}$      		& reflection of standards measured with test network direct  \\
$\mathbf{\Gamma}'_\text{R}$      		& reflection of standards measured with test network reverse \\
$\mathbf{S}_\text{RP}$           		& S-parameters that take reference plane to calibration  \\
$\mathbf{S}_\text{D}$            		& S-parameters of test network direct  \\
$\mathbf{S}_\text{R}$            		& S-parameters of test network reverse \\
\bottomrule
\end{tabular}
\end{table}

\begin{equation}
\text{FoM} = \sum_k (\Delta 1_k + \Delta 2_k + \Delta 3_k),
\label{equation_FoM}
\end{equation}

\noindent where

\begin{align}
\Delta 1 =& |S_{11\text{D}} - S_{11\text{R}}|,\label{equation_delta1}\\
\Delta 2 =& |S_{12\text{D}}S_{21\text{D}} - S_{12\text{R}}S_{21\text{R}}|,\label{equation_delta2}\\
\Delta 3 =& |S_{22\text{D}} - S_{22\text{R}}|,\label{equation_delta3}
\end{align}

\noindent and $\sum_k$ sums over frequency. This figure of merit quantifies in a single number the differences between the complex S-parameters in direct and reverse modes (subscripts D and R, respectively). Additionally, it does not require the separate use of $S_{12}$ and $S_{21}$ and therefore their multiplication as provided by (\ref{equation_s_parameters}) can be used directly in the $\Delta 2$ term.

An extra layer of correction is required by the D/R method. In (\ref{equation_gamma_shifted}$)-($\ref{equation_s_parameters}) it has been implicitly assumed that the values recorded at the reference plane (the primed quantities) are calibrated. In practice, this calibration will be as good as the assumptions used for the calibration standards measured at this plane. The direct and reverse S-parameters of the test network will have a chance of matching only if the reference plane is calibrated. To account for this aspect, the D/R method also requires measuring the standards at the reference plane.

The standards are assumed to have reflections modeled by $\mathbf{\Gamma}_\text{M}$, with parameters $\mathbf{p}$. The D/R method involves conducting three sets of measurements as follows: 

\begin{enumerate}
\item Measure calibration standards at reference plane, $\mathbf{\Gamma}'_\text{RP}$.
\item Connect test network in direct mode (port 1 facing reference plane).
\item Measure standards at port 2 of test network, $\mathbf{\Gamma}'_\text{D}$.
\item Connect test network in reverse mode (port 2 facing reference plane).
\item Measure standards at port 1 of test network, $\mathbf{\Gamma}'_\text{R}$.
\end{enumerate}

The nomenclature is summarized in Table \ref{table_nomenclature}, where the vector quantities $\mathbf{\Gamma}$ represent the reflection coefficients of the open, short, and load standards, and the $\mathbf{S}$ quantities represent four-element S-parameter matrices. Both have an implicit dependence on frequency. Fig. \ref{figure_diagram} depicts the three sets of measurements required by the method.

With the measurements at hand and for a vector of estimates $\mathbf{p}_i$, the FoM is evaluated as follows:

\begin{figure}[t]
\centering
\includegraphics[width=0.42\textwidth]{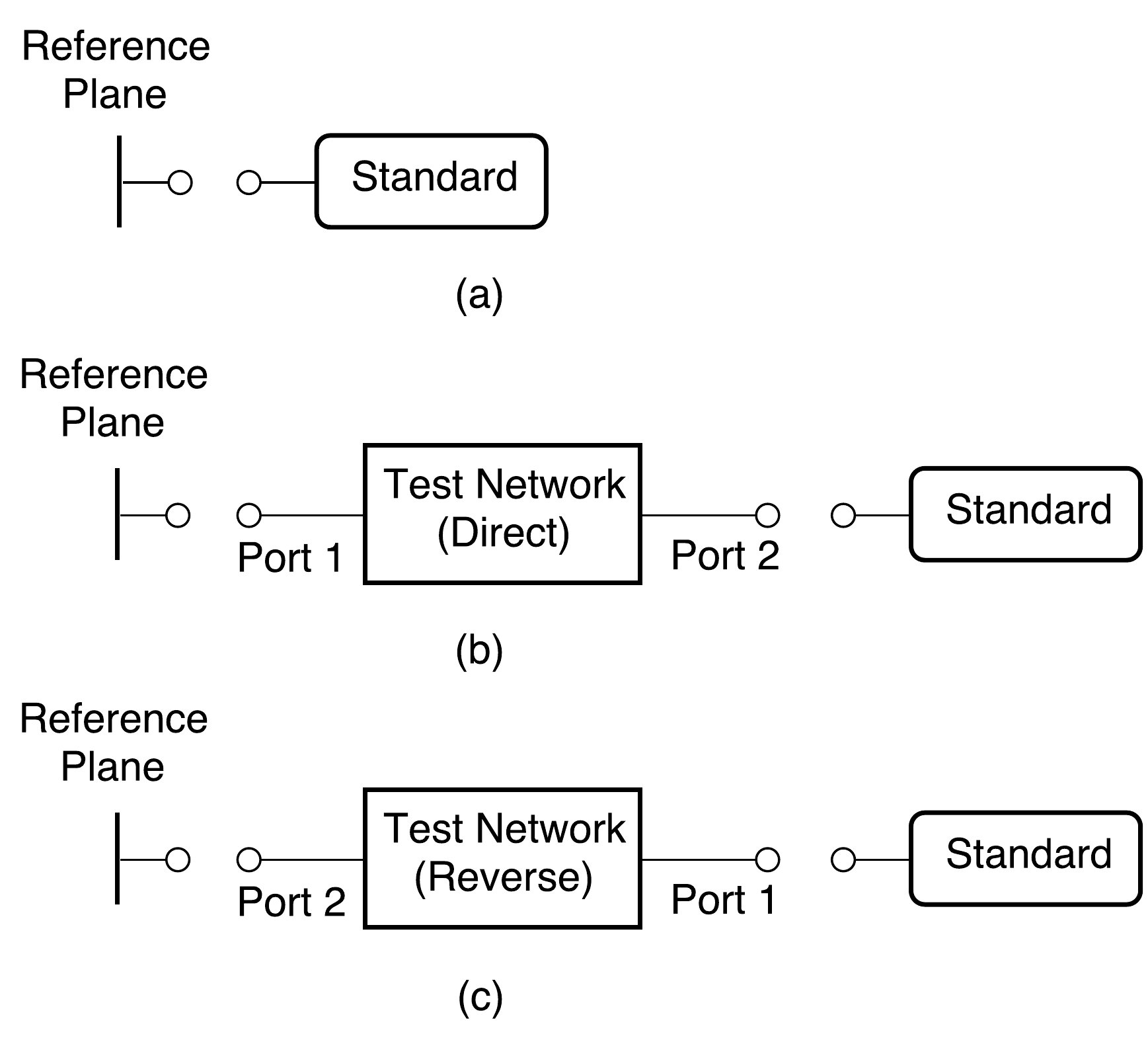}
\caption{Measurements involved in the D/R method. The SOL standards are measured (a) at the reference plane, (b) at the end of the test network in direct mode, and (c) at the end of the test network in reverse mode.}
\label{figure_diagram}
\end{figure}

\begin{enumerate}
\item Compute $\mathbf{S}_{\text{RP}i}$ using (\ref{equation_s_parameters}), where $\mathbf{\Gamma}_{\text{M}i}$ represents the assumed values for the standards evaluated at $\mathbf{p}_i$, and $\mathbf{\Gamma}'_\text{RP}$ represents their measurement at the reference plane.
\item De-embed $\mathbf{S}_{\text{RP}i}$ from the measurements $\mathbf{\Gamma}'_\text{D}$ and $\mathbf{\Gamma}'_\text{R}$ using (\ref{equation_gamma_actual}). The new quantities are labeled $\mathbf{\Gamma}'_{\text{D}i}$ and $\mathbf{\Gamma}'_{\text{R}i}$.
\item Compute $\mathbf{S}_{\text{D}i}$ and $\mathbf{S}_{\text{R}i}$ using (\ref{equation_s_parameters}), where $\mathbf{\Gamma}_{\text{M}i}$ represents the assumed values for the standards evaluated at $\mathbf{p}_i$, and $\mathbf{\Gamma}'_{\text{D}i}$ and $\mathbf{\Gamma}'_{\text{R}i}$ represent the measurements of the standards through the test network in direct and reverse mode after de-embedding $\mathbf{S}_{\text{RP}i}$.
\item Evaluate FoM$_i$ using (\ref{equation_FoM}).
\end{enumerate}

The minimum FoM in parameter space can be found using approaches such as grid search or iterative algorithms. The following sections provide implementation details for the cases presented in this work.

\section{Simulations}
\label{section_simulations}

Simulations are performed to demonstrate the method and understand its capabilities and limitations. The example in this section involves the simultaneous estimation of free parameters of the model for coaxial standards presented in the appendix. Three parameters are estimated: 1) the offset loss of the short, 2) the offset delay of the load, and 3) the offset loss of the load. They are assigned nominal values of $2.4$ G$\Omega$s$^{-1}$, $30$ ps, and $2.3$ G$\Omega$s$^{-1}$ respectively. The other (fixed) parameters take the fiducial values of the Keysight 85033E standards.

Several alternatives were considered for the design of the test network. During the analyses it was found that the free parameters are estimated with the lowest uncertainty when the difference between $S_{11}$ and $S_{22}$ is maximized, while keeping $|S_{12}|$ $=$ $|S_{21}|$ as high as possible (close to 1). These conditions cannot be achieved simultaneously for a wide range of frequencies, and therefore some compromises have to be made considering practical aspects. The chosen network consists of a circuit with a capacitor between the two ports and an inductor between port 2 and ground. This network is easy to implement and its performance can be optimized at a specific frequency while remaining useful in a wider range.

The S-parameters representing the test network are related to the impedances of the capacitor ($Z_\text{C}$) and inductor ($Z_\text{L}$) by

\begin{align}
S_{11} &= \frac{Z_\text{C} Z_\text{L} + Z_\text{C} Z_0 - Z_0^2}{Z_\text{C} Z_\text{L} + Z_\text{C} Z_0 + 2Z_\text{L} Z_0 + Z_0^2},\\
S_{22} &= \frac{Z_\text{C} Z_\text{L} - Z_\text{C} Z_0 - Z_0^2}{Z_\text{C} Z_\text{L} + Z_\text{C} Z_0 + 2Z_\text{L} Z_0 + Z_0^2},\\
S_{12} &= S_{21} = \frac{2Z_\text{L} Z_0}{Z_\text{C} Z_\text{L} + Z_\text{C} Z_0 + 2Z_\text{L} Z_0 + Z_0^2},
\label{equation_L_circuit}
\end{align}

\noindent where $Z_0 =$ $50$ $\Omega$. The capacitance and inductance are chosen so that the S-parameters enable a precise estimation of the free parameters. This is discussed in more detail in section \ref{section_measurements} in the context of estimation from actual measurements. Values of $5$ pF and $17$ nH are used in this simulation because they provide near-optimum performance.

Synthetic noisy data are produced to represent the three sets of measurements. A $1\sigma$ noise level of $1\times 10^{-4}$ (linear) is assigned to the real and imaginary parts of the synthetic measurements. This value is realistic for the VNA settings used during actual measurements.

\begin{figure}[t!]
\centering
\includegraphics[width=0.44\textwidth]{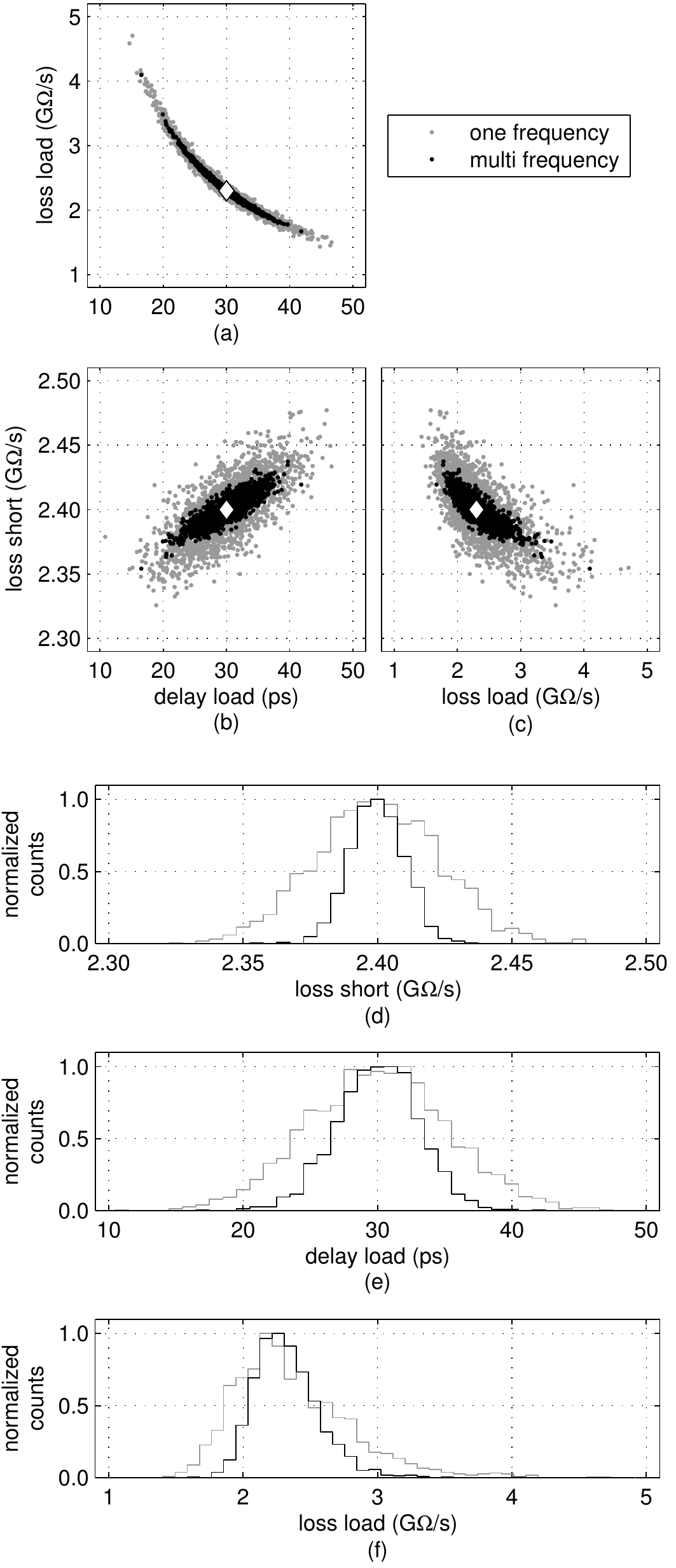}
\caption{Summary of the simulations conducted to demonstrate the D/R method assuming coaxial standards modeled as described in the appendix. The free parameters are the offset loss of the short, the offset delay of the load, and the offset loss of the load. The (a)-(c) panels show the input values of the free parameters as white diamonds and the two-dimensional distributions recovered from the simulations due to measurement noise. The (d)-(f) panels show the marginalized distributions. Two cases are simulated. In the first case (gray), only a measurement at $1000$ MHz is used to estimate the parameters. The second case (black) uses measurements in the range between $50$ and $1000$ MHz with step of $50$ MHz. The same two-port network and noise are used in both cases. The second case produces lower uncertainties in the estimates.}
\label{figure_simulation}
\end{figure}

The free parameters of the standards are estimated by following the recipe at the end of section \ref{section_method} and finding the minimum FoM through an iterative algorithm for unconstrained nonlinear optimization based on a quasi-Newton method, available in MATLAB as the \texttt{fminunc} function. This alternative is preferred over a direct grid search, which for three parameters is significantly more intensive computationally. The effect of measurement uncertainty on the estimates is determined by repeating this process for $N$ $=$ $2000$ realizations of noise. This number of repetitions keeps the standard deviation of the parameters stable to within $5$\%.

Two scenarios are explored. In the first one, the parameters are estimated only from measurements at $1000$ MHz, whereas the second case uses data between $50$ and $1000$ MHz in steps of $50$ MHz. This is done in order to make evident the benefits of conducting measurements in a broader range.

The results are summarized in Fig. \ref{figure_simulation}. The top plots present the covariance between parameters and the bottom plots show the marginalized distributions. The parameter estimated with the highest precision is the offset loss of the short, with a standard deviation of $0.023$ G$\Omega$s$^{-1}$ for a measurement at $1000$ MHz and $0.010$ G$\Omega$s$^{-1}$ for the broader measurement. The offset delay of the load has a standard deviation of $5.2$ ($3.0$) ps, and the offset loss of the load of $0.446$ ($0.241$) G$\Omega$s$^{-1}$ for the single (multi) frequency case. These two parameters are not as well constrained as the offset loss of the short due to their strong $\sim$ $1/x$ correlation, and therefore significant improvement is possible if one of them were kept fixed during estimation.

Although the simulation described focuses on estimating parameters of the standard offsets, the D/R method is equally applicable for improving the characterization of the termination elements. For example, using a similar strategy as above, simulations were performed to estimate the coefficients of the polynomials that model the capacitance and inductance of the open and short terminations, respectively (see (\ref{equation_capacitance_open}) and (\ref{equation_inductance_short})). The estimations were done separately, first for the open and then for the short, with four free parameters at a time. As expected, it was necessary to increase the frequency range to $9$ GHz (the highest allowed by this calibration kit) and to reduce the noise level below $1\times10^{-5}$ to be able to constrain the frequency dependence properly and estimate the parameters robustly and with precision. Table \ref{table_open_short_termination} presents the estimates for the polynomial coefficients using simulated measurements between $500$ MHz and $9$ GHz with a step size of $500$ MHz and noise of $1\times10^{-6}$. Clearly the precision of the estimation decreases as the degree of the polynomial term increases, especially for the inductance coefficients. This type of estimation would be challenging in practice, but this example shows the flexibility of the method when the measurement setup is properly optimized for a specific application.

\begin{table}\centering
\ra{1.3}
\caption{Estimation of Capacitance and Inductance Coefficients for the Open and Short Standards from Simulated Data}
\label{table_open_short_termination}
\begin{tabular}{@{}llll@{}}
\toprule
Coefficient & Input Value & Recovered Value ($\pm1\sigma$) & Units  \\
\midrule
$\hat{C}_0$ & $+4.943$  & $+4.94297\pm0.00094$ &   $\times10^{-14}$ (F) \\
$\hat{C}_1$ & $-3.101$  & $-3.099\pm0.082$      &   $\times10^{-25}$ (F Hz$^{-1}$) \\
$\hat{C}_2$ & $+2.317$  & $+2.31\pm0.20$       &   $\times10^{-35}$ (F Hz$^{-2}$) \\
$\hat{C}_3$ & $-1.597$ & $-1.6\pm1.4$      &   $\times10^{-46}$ (F Hz$^{-3}$) \\
\midrule
$\hat{L}_0$ & $+2.077$   & $+2.077\pm0.018$   &   $\times10^{-12}$ (H) \\
$\hat{L}_1$ & $-1.085$   & $-1.08\pm0.15$   &   $\times10^{-22}$ (H Hz$^{-1}$) \\
$\hat{L}_2$ & $+2.171$   & $+2.1\pm3.7$   &   $\times10^{-33}$ (H Hz$^{-2}$) \\
$\hat{L}_3$ & $-1.000$   & $-1\pm26$  &   $\times10^{-44}$ (H Hz$^{-3}$) \\
\bottomrule
\end{tabular}
\end{table}

\section{Measurements and Results}
\label{section_measurements}

Real measurements were conducted with the purpose of estimating the offset delay of the $50$-$\Omega$ load from a Keysight 85033E calibration kit modeled as described in the appendix. The other parameters of the kit take their nominal values, with the exception of the termination impedance of the load which takes the value of its DC resistance measured with a 6.5-digit precision ohmmeter. Specifically, this quantity is measured using a cable assembly with an SMA connector at its end. First, the resistance of the assembly itself is measured by connecting the short standard; then, the load is connected, and its resistance is obtained by subtracting from the reading the assembly resistance. Uncertainty is estimated at $4\times10^{-3}$ $\Omega$, dominated by fluctuations of the assembly resistance.

The calibration kit has $3.5$-mm connectors. This has implications for the gender of the connectors in the system, especially when considering that the test network has to be measured in direct and reverse modes. The connectors of the test network are female, and therefore:
\begin{enumerate}
\item the connector at the reference plane has to be male, 
\item the connector of the standards measured at the reference plane has to be female, and
\item the connector of the standards measured at the end of the test network has to be male. 
\end{enumerate}

\begin{figure}[t]
\centering
\includegraphics[width=0.5\textwidth]{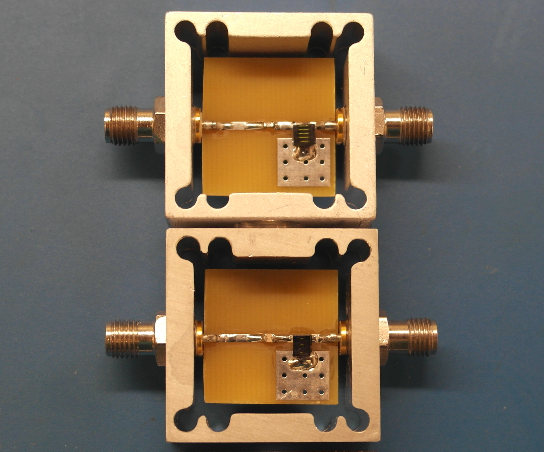}
\caption{Two-port networks 1 and 2, as implemented. They consist of lumped surface-mount elements on a double-layer FR4 board (ground layer on the bottom side). The PCBs are $20\times20$ mm$^2$. The connectors are female SMA and the metal enclosure has top and bottom covers, not shown here.}
\label{figure_networks}
\end{figure}

Thus, it is necessary to use the two sets of standards of opposite genders available in the calibration kit. There are no good alternatives to this arrangement, other than inverting all the genders. Given that the physical characteristics of the male and female $50$-$\Omega$ loads are almost identical, they are assumed to have the same offset delay. This is consistent with the other parameters of the calibration kit which are also almost identical between genders, and helps to keep the number of free parameters and uncertainties to a minimum. The measured DC resistances of the female and male loads are $49.995$ $\Omega$ and $50.010$ $\Omega$ respectively.

The topology chosen for the test network is a circuit consisting of a capacitor and an inductor, as described in section \ref{section_simulations}. The capacitance and inductance are chosen so that they minimize the uncertainty of the load delay at a specific frequency within the measurement range. With this approach it is possible to produce more than one network in the range of interest. This is useful for cross-checking and validating the estimates even if they do not have the highest precision.

Two networks are implemented. Network 1 is optimized at $600$ MHz and network 2 at $1000$ MHz. The optimization is conducted through simulations by sweeping over a range of values of capacitance and inductance until the combination that produces the lowest uncertainty in the load delay is found, for a given level of measurement noise. The values found for network 1 are $4.7$ pF and $17$ nH, and for network 2 they are $4$ pF and $8$ nH. The networks are implemented using lumped surface-mount capacitors and inductors, soldered on double-layer $20\times20$ mm$^{2}$ FR4 boards with a ground layer on the bottom side. The connectors are female SMA, and the enclosures are made out of aluminum including their top and bottom covers. The networks are shown in Fig. \ref{figure_networks}.

\begin{figure}[t]
\centering
\includegraphics[width=0.50\textwidth]{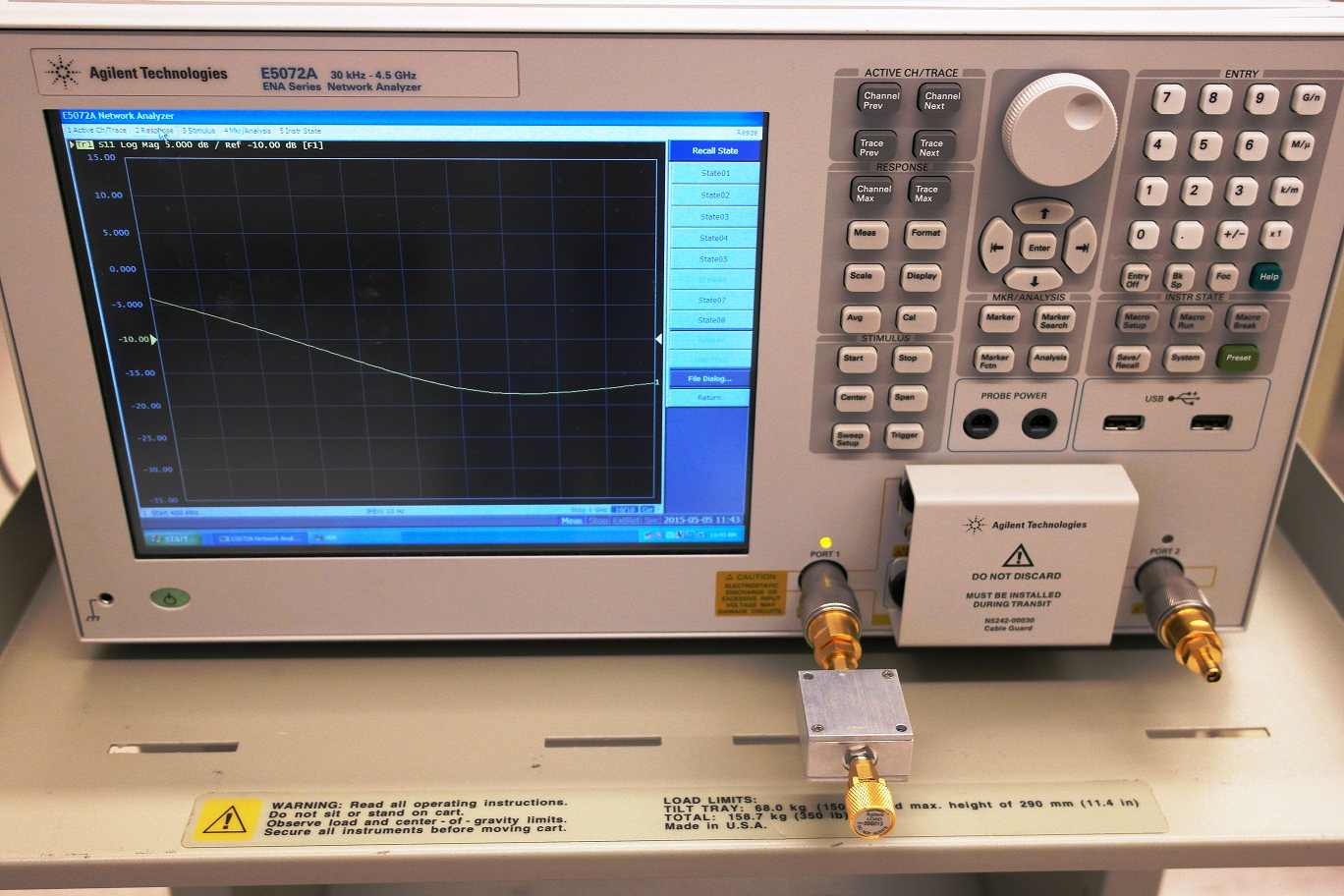}
\caption{Measurement setup. The picture shows the Keysight (Agilent) E5072A VNA, one of the test networks connected to the VNA port 1, and one of the calibration standards connected to the test network. The reference plane is at the male $3.5$-mm connector of the VNA port.}
\label{figure_setup}
\end{figure}

The three sets of measurements described in section \ref{section_method} were conducted with a Keysight E5072A VNA and the following settings: power of $0$ dBm, frequency between $400$ and $1000$ MHz in steps of $50$ MHz, bandwidth of $10$ Hz, and averaging of ten traces. The measurement of each standard at the reference plane and through the test network is repeated manually ten times, following a disconnection and reconnection. This is done in order to account for potential scatter due to limited connection repeatability, or instability in the performance of the test network or VNA. Fig. \ref{figure_setup} shows the measurement setup.

The sample average and standard deviation are computed for each repeated measurement. In particular, the largest scatter has a $1\sigma$ level of $5\times 10^{-4}$ and occurs for network 2. Most of this scatter has a systematic origin since the VNA settings result in $1\sigma$ noise below $1\times 10^{-4}$. The measurement uncertainty is modeled as Gaussian using the statistics computed from the repeated measurements. This uncertainty is propagated to the estimated load delay by processing $N$ $=$ $15000$ Monte Carlo realizations through the algorithm that identifies the lowest FoM.

Since there is only one free parameter, the minimum FoM is found by sweeping over values of load delays. The resolution of the sweep is $0.1$ ps, in the range between $-60$ and $60$ ps. The range extends toward negative values in order to confirm that the routine does not yield unphysical estimates.

\begin{figure}[t]
\centering
\includegraphics[width=0.48\textwidth]{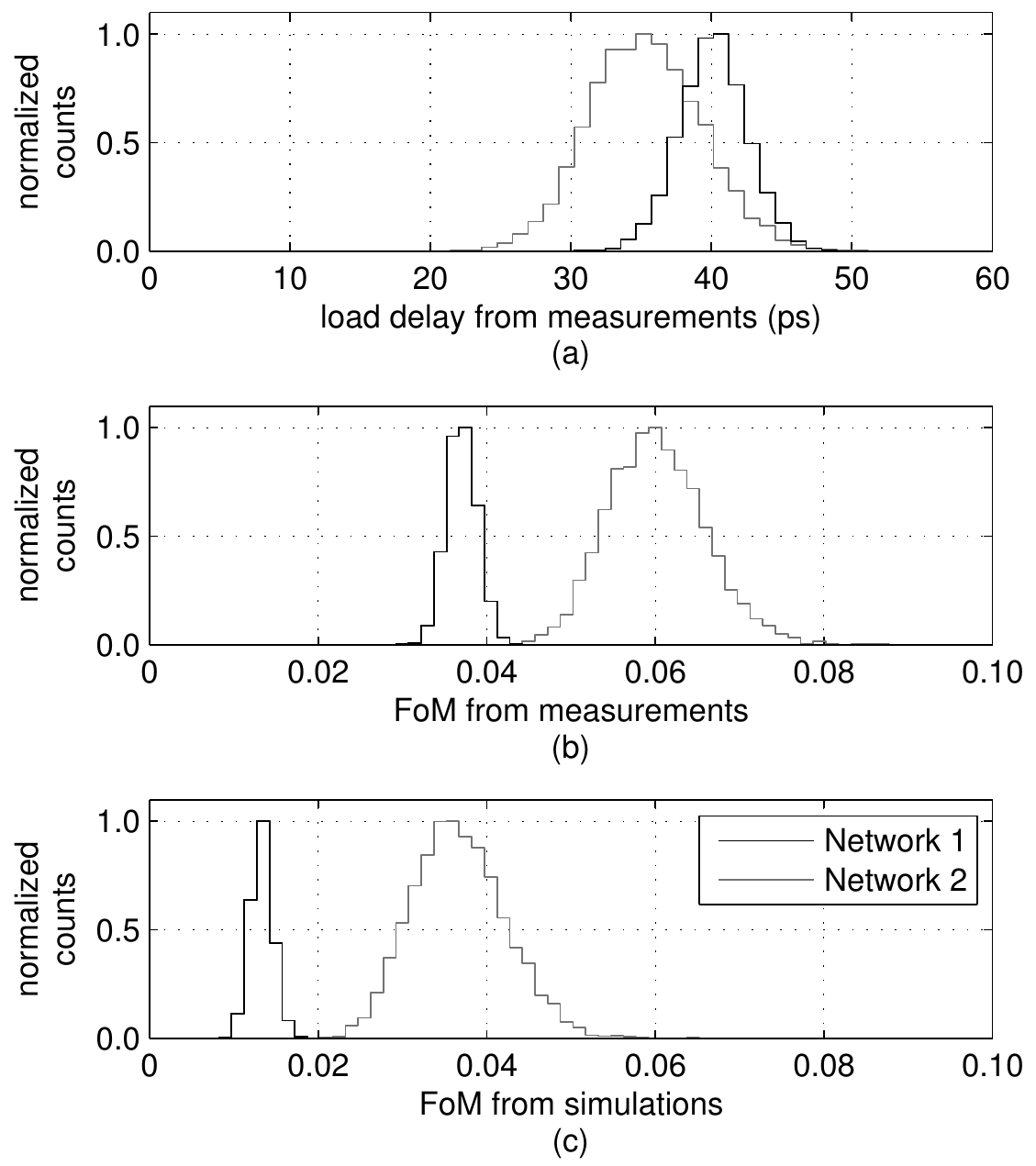}
\caption{Panel (a) shows distributions for the offset delay of the $50$-$\Omega$ load obtained from measurements through both test networks. The weighted average of the two distributions is $38.8\pm2.1$ ps ($1\sigma$ uncertainty). Panel (b) shows the figures of merit (FoM) corresponding to the distributions presented in panel (a). Panel (c) shows the FoM produced by a simulation with characteristics similar to the real case, but with perfect modeling of the standards and VNA. On average, they are lower than those from measurements by $0.024$.}
\label{figure_results}
\end{figure}

Fig. \ref{figure_results} presents the results of the estimation. The top panel shows the distributions of the load delay from measurements through both test networks. The averages and $1\sigma$ uncertainties are $40.1\pm2.4$ and $35.3\pm4.0$ ps for networks 1 and 2, respectively. The averages are different by $4.8$ ps with an uncertainty of $\sqrt{\sigma_{N1}^2 + \sigma_{N2}^2}=4.7$ ps, which corresponds to a $\sim$ $1\sigma$ significance. The poorer performance of network 2 can be attributed to its higher measurement scatter. Although the uncertainties are not optimal, the consistency of the estimates serves as verification of the method and its implementation. The definitive estimate is calculated as the weighted average of the two results, which yields $38.8\pm 2.1$ ps ($1\sigma$ uncertainty). 

The middle panel of Fig. \ref{figure_results} shows the distributions of the FoMs associated to the estimates of the load delay. The average FoMs are $0.037$ and $0.060$ for Network 1 and 2, respectively. In order to gain intuition about these results, a simulation was run in which noise, standards, VNA calibration, and test networks have realistic values. The two advantages of this simulation over the real case are: 1) the model used for the standards during the parameter estimation is correct, in the sense that it is the same as the one used to generate the synthetic data, and 2) the S-parameters of the networks in direct and reverse mode are identical, which is equivalent to assuming a perfectly linear VNA. The simulated FoMs are presented in the lower panel of the figure. Their most important feature when compared to the FoMs from measurements corresponds to averages which are lower by $0.024$ for both networks. The presence of excess residuals in the real case relative to the simulation is an indication that there are aspects of the measured setup which have not been modeled perfectly, such as the response of the standards or the performance of the VNA.

Due to the excess residuals in the FoMs, the estimate for the load delay reported in this work for this particular calibration kit can only be regarded as a first-order correction to the value provided by the manufacturer. Nonetheless, it still represents an improvement that helps mitigate inaccuracies in measurements of reflection coefficient and S-parameters. 

As future work, the D/R method could be improved by incorporating expectations from simulations into the optimization algorithm to help refine the measurement models and minimize systematics. Also, a broader range of test network topologies could be considered, aiming at selecting that with the highest sensitivity to errors in the parameter estimates.

\section{Verification}
\label{section_verification}
\begin{figure}
\centering
\includegraphics[width=0.5\textwidth]{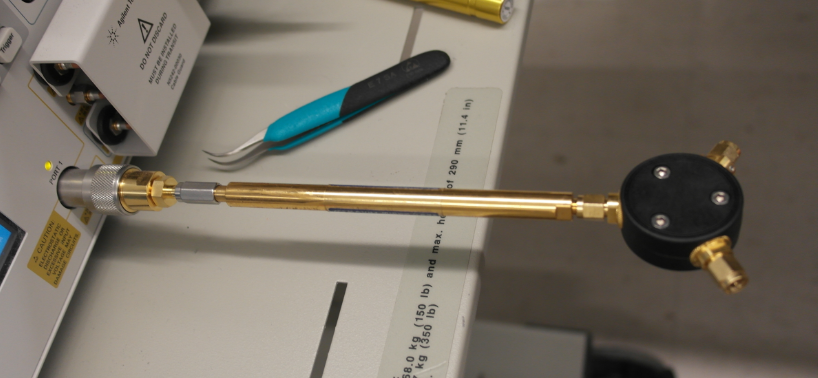}
\caption{Setup during measurements of the terminated airline. The black circular block at the end of the airline is a plastic holder for the three calibration standards.}
\label{figure_airline_measurement}
\end{figure}

We use an 8043S15 beadless airline from Maury Microwave ($15$-cm-long, $3.5$-mm connectors) to verify the new estimate for the load delay. The verification consists of comparing models for the reflection coefficient of the airline with measurements done after calibrating the VNA assuming both, the nominal value ($0$ ps) and the new estimate ($38.8$ ps) for the load delay. Five measurements are conducted at the reference plane: the open, short, and load calibration standards, and the airline terminated at the open and short standards. The setup during measurements is shown in Fig. \ref{figure_airline_measurement}.

\begin{figure}
\centering
\includegraphics[width=0.5\textwidth]{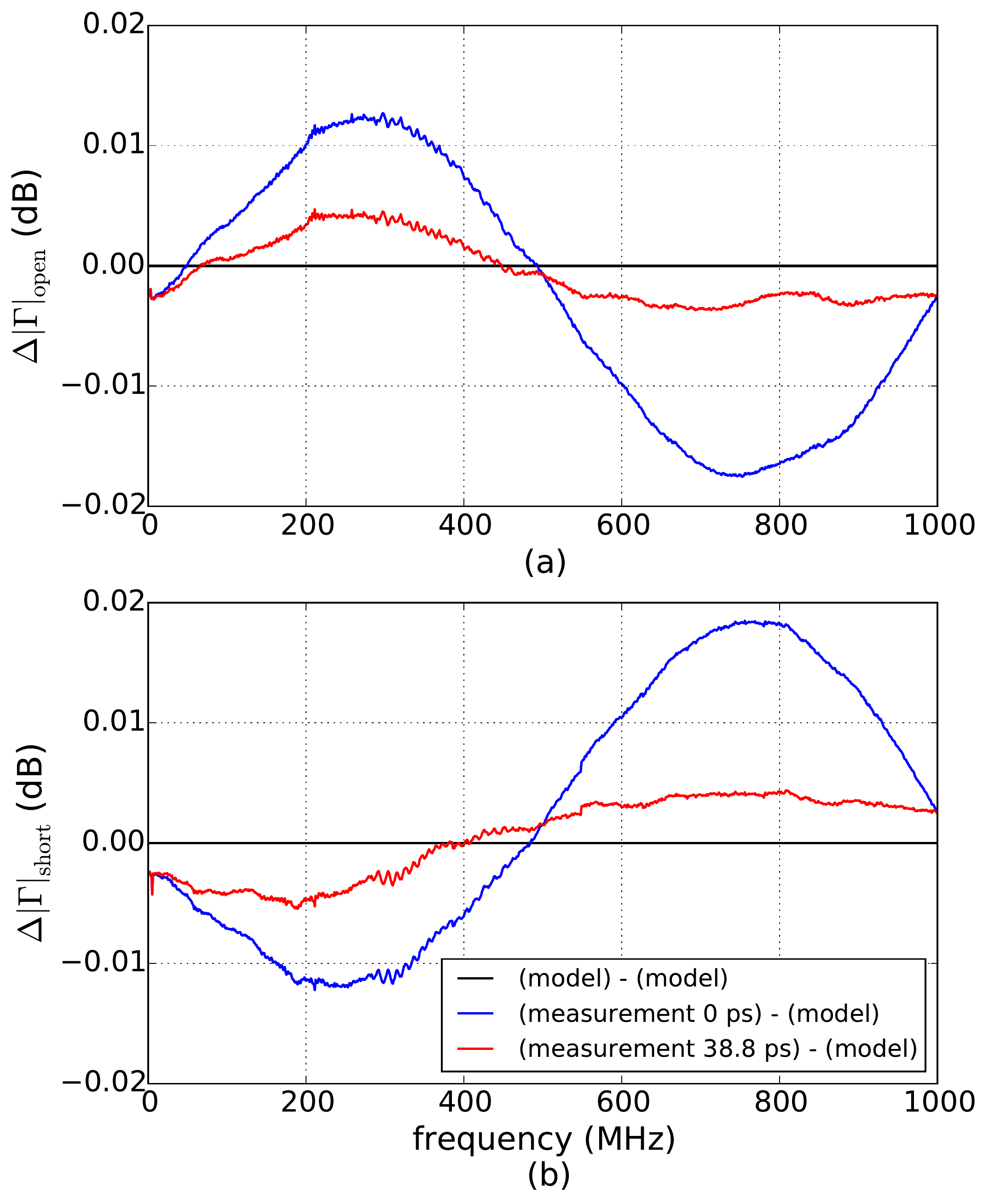}
\caption{Verification of the new estimate for the load delay using a $15$-cm airline terminated in the (a) open and (b) short standards. The blue line represents the difference in reflection magnitude between the terminated airline measured after calibrating the reference plane using $0$ ps for the load delay, and the model. The red line represents the difference between the measurement calibrated using $38.8$ ps  and the model. The black line corresponds to a perfect match to the model. The better agreement between the black and red lines verify that using $38.8$ ps for the load delay produces more accurate reflection measurements.}
\label{figure_verification}
\end{figure}

The terminated airline impedance is modeled as

\begin{equation}
Z_{\text{in}} = Z_{\text{char}} \frac{Z_{\text{ter}} + Z_{\text{char}} \tanh(\gamma \ell)} {Z_{\text{ter}}\tanh(\gamma \ell) + Z_{\text{char}}},
\end{equation}

\noindent where $\ell = 14.99$ cm is the electrical length of the airline and $Z_{\text{ter}}$ is the termination impedance, which in this case is given by (\ref{equation_impedance_open}) and (\ref{equation_impedance_short}). The characteristic impedance and propagation constant of the airline are given by

\begin{align}
Z_{\text{char}} &= \sqrt{ \frac{R + j\omega L} {G + j\omega C} }, \\
\gamma          &= \sqrt{ \left( R + j\omega L \right) \left( G + j\omega C\right) },
\end{align}

\noindent with distributed parameters defined as

\begin{align}
R &= \sqrt{\frac{\omega\mu_0}{2\sigma}}\left( \frac{1}{2\pi r_i} + \frac{1}{2\pi r_o} \right)\\
G &= 0 \\ 
C &= \frac{2 \pi \epsilon_{\text{air}}}{\ln\left(\frac{r_o}{r_i}\right)}\\
L &= 2L_{\text{cond}} + L_{\text{dielec}} = 2     \sqrt{\frac{2\mu_0}{\omega\sigma}} \frac{1}{4\pi r_i} + \frac{\mu_0}{2\pi}\ln\left(\frac{r_o}{r_i}\right).
\end{align}

These expressions represent an approximation to the full theory of \cite{ramo} valid for a large ratio of conductor radius to skin depth. The inductances-per-unit-length $L_{\text{cond}}$ and $L_{\text{dielec}}$ correspond to the conductor and air dielectric of the airline, respectively. In all these expressions the fundamental quantities are the angular frequency, $\omega$, the permeability of vacuum $\mu_0$, the permittivity of air, $\epsilon_{\text{air}}$, and the conductivity of the airline conductor $\sigma$. The outer radius of the inner conductor is $r_i=0.7595$ mm and the inner radius of the outer conductor is $r_o=1.7501$ mm.

The center and outer conductors of the airline are made out of beryllium copper and plated with copper and gold. The plating thicknesses are up to $0.25$ $\mu$m and $0.5$ $\mu$m, respectively. This is signifficantly lower than the corresponding skin depth, which for copper (gold) is $6.6$ ($7.9$) $\mu$m  at $100$ MHz and $2.1$ ($2.5$) $\mu$m at $1$ GHz. Thus, the airline conductivity primarily corresponds to that of beryllium copper and is obtained through four-wire measurements of the airline resistivity, resulting in a conductivity of $16.5\pm1.5\%$ relative to copper.

Figure \ref{figure_verification} presents the results of the verification. The top (a) and bottom (b) panels of the figure correspond to the airline terminated in the open and short standards, respectively. Both panels show in blue the difference between the reflection measurements calibrated using $0$ ps for the load delay, and the model. The red lines represent the difference between the measurement using $38.8$ ps for the load delay and the model. The black line corresponds to a reference for the case of perfect match to the model. The better agreement between the black and red lines indicates that a higher measurement accuracy is achieved when calibrating the reference plane using a value of $38.8$ ps for the load delay, as estimated with the D/R method.

To quantify the improvement we compute the RMS between the magnitude (in linear units) of the measurement calibrated with each delay value, and the model. The results are $12.0\;(12.6) \times 10^{-4}$ using $0$ ps and $3.1\;(3.8) \times 10^{-4}$ using $38.8$ ps for the airline terminated in the open (short) standard. A lower RMS for both terminations verify that a load delay of $38.8$ ps produces more accurate reflection measurements.

Although the measurements do not match the model perfectly, the results of this verification are robust against realistic uncertainties in model parameters such as the mechanical dimensions and conductivity of the airline. Better match and stronger verification could be achieved at these low frequencies using 1) a longer airline, to produce more ripples over frequency, and 2) conductors without plating, for a direct determination of their conductivity.

\section{Conclusion}
\label{section_conclusion}
This work introduced the one-port direct/reverse method for improving the characterization of VNA reflection standards. The method was demonstrated through simulations and used to estimate the offset delay of the $50$-$\Omega$ load from a Keysight 85033E $3.5$-mm calibration kit. For practical reasons, the male and female loads had to be measured during the procedure and it was assumed that both had the same delay.

Measurements using two different test networks optimized for different frequencies produced consistent results for the delay, with a weighted average of $38.8\pm2.1$ ps ($1\sigma$ uncertainty). This result was verified measuring the reflection coefficient of a $15$-cm beadless airline terminated in an open and short standard after calibrating the reference plane using $0$ and $38.8$ ps for the load delay. The measurements calibrated with $38.8$ ps agree better with theoretical models for the terminated airline. 

Future work could involve maximizing the sensitivity of the measurements to parameter errors by selecting the test network from a broad range of topologies.

\appendix

The reflection coefficient seen at the input of the open, short, and load coaxial VNA standards is modeled as lumped termination elements at the end of transmission line sections, or offsets. This represents an approximation to transmission line theory \cite{ramo}, \cite{marks}, and has been presented in \cite{degroot} and \cite{scott}. In this model, the reflection coefficient of the standards is given by

\begin{equation}
\Gamma=\frac{\Gamma_{\text{off}}\left(1-e^{-2\gamma\ell}-\Gamma_{\text{off}}\Gamma_{\text{ter}}\right) + e^{-2\gamma\ell} \Gamma_\text{ter}}{1-\Gamma_{\text{off}}\left[e^{-2\gamma\ell} \Gamma_{\text{off}} +\Gamma_{\text{ter}}\left(1-e^{-2\gamma\ell}\right)\right]},
\label{equation_standards}
\end{equation}

\begin{equation}
\Gamma_{\text{off}}=\frac{Z_{\text{off}}-50}{Z_{\text{off}}+50},\;\;\;\Gamma_{\text{ter}}=\frac{Z_{\text{ter}}-50}{Z_{\text{ter}}+50},
\end{equation}

\noindent where:

\begin{itemize}
\item $Z_{\text{ter}}:$ impedance of termination.
\item $Z_{\text{off}}:$ lossy characteristic impedance of offset.
\item $\ell$: length of offset.
\item $\gamma$: propagation constant of offset.
\end{itemize}

The offsets are described in terms of their one-way loss evaluated at $1$ GHz ($\delta_{1\text{GHz}}$), one-way delay ($\tau$), and characteristic impedance assuming no loss ($Z_0$). Under the realistic assumption of zero conductance ($G=0$) in the distributed parameter model of transmission lines, and after a first order approximation, the lossy characteristic impedance and the propagation constant of the offsets can be expressed in terms of the previous quantities as

\begin{align}
Z_{\text{off}}&=Z_0 +\left(1 - j\right)\left(\frac{\delta_{1\text{GHz}}}{4\pi f}\right)\sqrt{\frac{f}{10^{9}}},\\
\gamma\ell&=j 2\pi f\tau + \left(1 + j\right) \left(\frac{\tau \delta_{1\text{GHz}}}{2 Z_0}\right)\sqrt{\frac{f}{10^9}},
\end{align}

\noindent where $f$ represents frequency in hertz.

The impedance of the terminations of the open and short is given by

\begin{align}
Z_{\text{ter},\;\text{open}}  & = \frac{-j}{2\pi f \cdot C_{\text{open}}},\label{equation_impedance_open}\\
Z_{\text{ter},\;\text{short}} & = j2\pi f \cdot L_{\text{short}},\label{equation_impedance_short}
\end{align}

\noindent with

\begin{align}
C_{\text{open}}  & = \hat{C}_0 + \hat{C}_1f + \hat{C}_2f^2 + \hat{C}_3f^3,\label{equation_capacitance_open}\\
L_{\text{short}} & = \hat{L}_0 + \hat{L}_1f + \hat{L}_2f^2 + \hat{L}_3f^3.\label{equation_inductance_short}
\end{align}

The $\hat{C}$ and $\hat{L}$ quantities are the coefficients of the third-degree frequency-dependent polynomials that model the capacitance and inductance, respectively. 

The termination impedance of the load is usually assumed to be real and equal to $Z_0$, i.e., $50$ $\Omega$. However, in this work it takes the value of its DC resistance measured with a precision ohmmeter, as suggested in \cite{blackham} and \cite{ridler}.

\section*{Acknowledgment}
We would like to thank Hamdi Mani for his assistance with measurements at the ASU Low-frequency Cosmology Laboratory.

\ifCLASSOPTIONcaptionsoff
  \newpage
\fi


\newpage

\begin{IEEEbiographynophoto}{Raul A. Monsalve}
(M'2014) obtained his Ph.D in Physics in 2012 from the University of Miami, Florida, working on measuring the polarization of the cosmic microwave background radiation. In 2007 he obtained a B.S. in Electronics Engineering from the University of Concepcion, Chile, with thesis work on radio-frequency instrumentation. Currently, he works as a Research Associate at Arizona State University and the University of Colorado Boulder on experimental efforts to detect the redshifted 21 cm spectral line emitted by neutral Hydrogen in the early Universe. Dr. Monsalve is a member of the IEEE, the American Astronomical Society (AAS), and the American Physical Society (APS).
\end{IEEEbiographynophoto}

\begin{IEEEbiographynophoto}{Alan E. E. Rogers}
(M'71) received the Ph.D. in 1967 in Electrical Engineering from the Massachusetts Institute of Technology (MIT). He joined the staff of MIT Haystack Observatory in 1968 where he carried out research in Radio and Radar Interferometry. He aided in the development of Very Long Baseline Interferometry (VLBI) for Geodesy and Astronomy. From 1994 to 2002 he worked with industry in the development of radio location systems for cellular phones. He retired in 2006 to his current position of Research Affiliate at Haystack Observatory. His current interests include radio arrays and spectrometers specializing in the detection and measurement of weak radio astronomy signals requiring very long integration times and accurate calibration. Dr. Rogers is a member of the IEEE, the American Geophysical Union (AGU), the American Astronomical Society (AAS), and the American Association for the Advancement of Science (AAAS).
\end{IEEEbiographynophoto}

\begin{IEEEbiographynophoto}{Thomas J. Mozdzen} 
(M'78) received the B.S. in 1978 in Physics and the M.S. in 1980 in Electrical Engineering from the University of Illinois at Urbana-Champaign, and the M.S. in 1985 in Physics from the University of Texas at Dallas. He joined Mostek Corporation in 1980 as a reliability physics engineer. He worked as a digital circuit design engineer at Mostek Corp., Siemens GmbH, and Intel Corp. until 2008. He is currently working on his Ph.D. in Astrophysics at Arizona State University in the area of the Epoch of Reionization detection via the global redshifted 21 cm signal. He holds 17 U.S. patents and is a member of the IEEE and the American Physical Society (APS).
\end{IEEEbiographynophoto}

\begin{IEEEbiographynophoto}{Judd D. Bowman}
(M'2014) received his Ph.D. in 2007 in Physics from the Massachusetts Institute of Technology (MIT). In 2010, after completing a Hubble Fellowship at Caltech, he joined the faculty of Arizona State University’s School of Earth and Space Exploration, where he is now an associate professor. His research interests include precision radio astronomy instrumentation, the astrophysics of the early Universe, and technologies for formal and informal STEM learning. Prof. Bowman has aided in the design and management of the Murchison Widefield Array (MWA) and the Hydrogen Epoch of Reionization Array (HERA) radio telescopes.  He is Principal Investigator of the Experiment to Detect the Global EoR Signal (EDGES). Prof. Bowman is a member of the IEEE, the International Union for Radio Science (URSI), the American Astronomical Society (AAS), and the American Association for the Advancement of Science (AAAS).
\end{IEEEbiographynophoto}


\begin{thebibliography}{1}




\bibitem{wong2}
K. H. Wong, ``Characterization of Calibration Standards by Physical Measurements,'' \emph{39th ARFTG Conf. Dig. Spring}, pp. 53-62, June 1992. 


\bibitem{kirby}
P. Kirby, L. Dunleavy, and T. Weller, ``Load Models for CPW and Microstrip SOLT Standards on GaAs,'' \emph{56th ARFTG Conf. Dig. Fall}, pp. 1-11, November 2000. 


\bibitem{walker}
D. K. Walker, D. F. Williams, and J. M. Morgan, ``Planar Resistors for Probe Station Calibration,'' \emph{40th ARFTG Conf. Dig. Fall}, pp. 1-9, December 1992.


\bibitem{eio}
C. P. Ei{\o}, S. J. Protheroe, and N. M. Ridler, ``Characterising Beadless Air Lines as Reference Artefacts for S-parameter Measurements at RF and Microwave Frequencies,'' \emph{IEE Proc. Sci. Meas. Technol.}, vol. 153, no. 6, pp. 229-234, November 2006.



\bibitem{degroot}
D. C. DeGroot, K. L.Reed, and J. A. Jargon, ``Equivalent Circuit Models For Coaxial OSLT Standards,'' \emph{54th ARFTG Conf. Dig. Spring}, pp. 1-13, December 2000.


\bibitem{scott}
J. B. Scott, ``Investigation of a Method to Improve VNA Calibration in Planar Dispersive Media Through Adding an Asymmetrical Reciprocal Device,'' \emph{IEEE Transactions on Microwave Theory and Techniques}, vol. 53, no. 9, pp. 3007-3013, September 2005.




\bibitem{ramo}
S. Ramo and J. B. Whinnery, \emph{Fields and Waves in Modern Radio}, 2nd ed., New York: John Wiley \& Sons, 1953.



\bibitem{marks}
R. B. Marks and D. F. Williams, ``A General Waveguide Circuit Theory,'' \emph{Journal of Research of the National Institute of Standards and Technology}, vol. 97, no. 5, pp. 533-562, October 1992.







\bibitem{rytting}
D. Rytting, ``Network Analyzer Error Models and Calibration Methods. RF 8: Microwave Measurements for Wireless Applications,'' ARFTG/NIST Short Course Notes, 1996.



\bibitem{ferrero1}
A. Ferrero, and U. Pisani, ``Two-Port Network Analyzer Calibration Using an Unknown ``\emph{Thru}'','' \emph{IEEE Microwave and Guided Wave Letters}, vol. 2, no. 12, pp. 505-507, December 1992.



\bibitem{ferrero2}
A. Ferrero, U. Pisani, and F. Sanpietro, ``Save the `Thru' in the A.N.A. Calibration,'' \emph{40th ARFTG Conf. Dig. Fall}, pp. 128-135, December 1992.


\bibitem{na}
Z. Na, Z. G. Hua, C. Ting, and L. Jie, ``Study on the Unknown Thru Calibration Technique,'' \emph{XXXIth URSI GASS}, pp. 1-4, August 2014. 



\bibitem{wong1}
K. Wong, ``The ``Unknown Thru'' Calibration Advantage,'' \emph{63rd ARFTG Conf. Dig. Spring}, pp. 73-81, June 2004.



\bibitem{blackham}
D. Blackham, and K. Wong, ``Latest Advances in VNA Accuracy Enhancements,'' \emph{Microwave Journal}, vol. 48, no. 7, pp. 78-94, July 2005.






\bibitem{jamneala1}
T. Jamneala, and M. Voo, ``Precision Calibration Coefficients for the Reciprocal Procedure,'' \emph{IEEE Microwave and Wireless Components Letters}, vol. 17, no. 5, pp. 406-408, May 2007.



\bibitem{jamneala2}
T. Jamneala, D. A. Feld, D. Blackham, K. H. Wong, and B. Zaini, ``Why Reciprocal Procedure Works?,'' \emph{Proc. RFIC Symposium}, vol. 1, p. 5, June 2006.  



\bibitem{ridler}
N. M. Ridler, and N. Nazoa, ``Using Simple Calibration Load Models to Improve Accuracy of Vector Network Analyzer Measurements,'' \emph{67th ARFTG Conf.}, pp. 104-110, June 2006. 



\bibitem{roberts}
T. Roberts, and J. Martens, ``Characterizing Calibration Standards Using One Airline as a Transfer Standard,'' \emph{ARFTG Conf.}, pp. 1-8, June 2014.






\bibitem{bowman}
J. D. Bowman, and A. E. E. Rogers, ``A Lower Limit of $\Delta$z $>$ 0.06 for the Duration of the Reionization Epoch,'' \emph{Nature}, vol. 468, pp. 796-798, December 2010.





\bibitem{rogers}
A. E. E. Rogers, and J. D. Bowman, ``Absolute Calibration of a Wideband Antenna and Spectrometer for Accurate Sky Noise Temperature Measurements,'' \emph{Radio Science}, vol. 47, RS0K06, July 2012.











\end{thebibliography}
\end{document}